\documentclass{article}

\usepackage{graphicx}
\usepackage{dcolumn}
\usepackage{bm}
\usepackage[mathlines]{lineno}
\usepackage[noblocks]{authblk}

\begin{document}

\title{Spin exchange collision mixing of the K and Rb ac Stark shifts}

\author{Yao Chen}
\author{Yan Lu}
\author{Jiancheng Fang}
\affil{\small
School of Instrumentation Science and Opto-electronics Engineering, Beihang University}

\date{\today}
\maketitle
\begin{abstract}
In a hybrid pumping alkali vapor cell that both K and Rb are filled, K atom spins are optically pumped by laser and Rb atom spins are polarized by the K spins through spin exchange. We find that the AC Stark shift of the Rb atoms is composed of not only the AC Stark shift of the Rb atoms caused by the far off resonant pumping laser which is tuned to the K absorption lines, but also the AC Stark shift of the K atom spins. The mixing of the light shifts through fast spin exchange between K and Rb atoms are studied in this paper and we demonstrate a K-Rb-$^{21}$Ne co-magnetometer in which the AC Stark shift of the Rb atoms are reduced by the collision mixing.
\end{abstract}

\maketitle


\section{Introduction}
When off-resonant, circularly-polarized light propagates through an atomic vapor cell, the vapor can experiences a virtual magnetic field which points the propagating direction of the laser\cite{happer1967}. With the development of hybrid optical pumping, the atomic vapor cell is filled with two or more spin species\cite{romalis2010}. One of the spin species is directly pumped by laser beam and others are polarized by spin exchange collision. In an atomic co-magnetometer based on K-Rb-$^{21}$Ne, the 770nm pump laser which is used to directly pump the K atom spins causes the ac Stark shift of the Rb spins at D2 line\cite{smiciklas2010}. We find that the ac Stark shift of K atoms could  mixes together with the Rb shift through spin exchange collision\cite{allred2002}. The spin exchange interaction between pair of alkali metal is so large that the pair of alkali metal exchange their spins rapidly.

Ac Stark shift is always a systematic error in atomic sensors. For example, in an atomic co-magnetometer based on K-Rb-$^{21}$Ne, the pumping beam ac Stark shift could cause the coupling of the input angular velocities of the x and y axis\cite{smiciklas2010}. For direct optical pumping that only one species alkali metal is utilized, the ac Stark shift can be suppressed by tuning the frequency of laser beam to the absorption peaks\cite{fangjianchen2012,kornack2005}. A diffusion suppression  method is also utilized to reduce the ac Stark shift\cite{sulai2013}. However, in an atomic sensors based on hybrid pumping, the ac stark shift could not be suppressed by the two methods. We find that collision mixing of the K and Rb ac Stark shifts which can be utilized to suppress the total ac stark shifts in an atomic sensors based on hybrid pumping. We can control the ac Stark shift of K to counteract the shift of Rb to reduce the total light shift.

We establish a K-Rb-$^{21}$Ne co-magnetometer to study the collision mixing of the ac Stark shifts of K and Rb atoms. The full Bloch equations\cite{hasegawa1959} of the three atom species including the K and Rb ac Stark shifts are established and solved analytic solutions are derived. The mixing of the K and Rb ac Stark shifts is experimentally studied.
\section{THEORY}
This study is based on an K-Rb-$^{21}$Ne  atomic co-magnetometer\cite{fangjiancheng2016}. The parameters in the model is based on the co-magnetometer. Suppose that in an alkali vapor cell which is filled with K, Rb and $^{21}$Ne gas atoms, the density ratio of K to Rb is $D_r=n_K/n_{Rb}$ which is on the order of 0.01. $n_K$ and $n_{Rb}$ are respectively the densities of K and Rb. K atoms are directly pumped by the D1 line laser and Rb atoms are pumped through spin exchange with K atoms. If the frequency of the laser beam $\nu$ is tuned away from the absorption center of the pressure\cite{Romalis1997} and power broadened\cite{Citron1977} D1 absorption line, there will be ac stark shift of K $L_{z}^{K}$in the light propagation direction. We denote the light propagation direction is the z direction. For Rb atoms, the far off-resonant laser will cause ac Stark shift $L_{z}^{Rb}$ on the D2 line of Rb atoms. We will show that the two ac Stark shifts will mix together through the spin exchange interaction.

Rb atoms are polarized by K atoms through spin exchange interaction. The spin exchange interaction conserve the total angular momentum and exchange the direction of the K and Rb spins. The spin transfer rate of Rb to K is given by $R_{Rb-K}^{SE}=n_{Rb}\sigma^{SE}\overline{v}_{Rb-K}$, where $\sigma^{SE}$ is the spin exchange cross section between Rb and K\cite{Cole1985}, $\overline{v}_{Rb-K}$ is the relative velocity between Rb and K. We can calculate the spin transfer rate of K to Rb to be $R_{K-Rb}^{SE}=D_r R_{Rb-K}^{SE}$. In a K-Rb-$^{21}$Ne co-magnetometer, the temperature of the vapor cell is about 473K and the density of the optically thick Rb atoms is about 5$\times$10$^{14}$/cm$^3$. The spin transfer rate of Rb to K is calculated to be approximately 1$\times$10$^{6}$sec$^{-1}$. The Rb and K atoms are in rapid spin exchange that the ac Stark shifts of them mix together.

The mixing of the ac Stark shifts is tested by the co-magnetometer. The Bloch equations which are utilized to model the co-magnetometer have been developed to describe the K-$^3$He co-magnetometer\cite{kornack2005}. In order to model the mixing of the ac Stark shifts, we add the equation which is utilized to describe K to the equation group. The full Bloch equations are given by:

\begin{eqnarray}\label{equation1}
\frac{\partial \bm{P^e_K}}{\partial t}=&&\frac{\gamma^e}{Q(P^e_{K-Rb})}(\bm{B}+\lambda_{K-Ne}M_0^n\bm{P^n}+\bm{L^K})\times \bm{P^e_K}
\nonumber\\
&&+\bm{\Omega}\times \bm{P^e_K} +(R_p\bm{s_p}-R_p\bm{P^e_K})/Q(P^e_{K-Rb})
\nonumber\\
&&+R_{Rb-K}^{SE}(\bm{P^e_{Rb}}-\bm{P^e_K})/Q(P^e_{K-Rb})
\nonumber\\
\frac{\partial \bm{P^e_{Rb}}}{\partial t}=&&\frac{\gamma^e}{Q(P^e_{K-Rb})}(\bm{B}+\lambda_{Rb-Ne}M_0^n\bm{P^n}+\bm{L^{Rb}})\times \bm{P^e_{Rb}}
\nonumber\\
&&-(R_{sd}+Q(P^e_{K-Rb})\frac{1}{T_2^{SE}})\bm{P^e_{Rb}}/Q(P^e_{K-Rb})\nonumber\\
&&+\bm{\Omega}\times \bm{P^e_{Rb}}+D_r R_{Rb-K}^{SE}(\bm{P^e_{K}}-\bm{P^e_{Rb}})/Q(P^e_{K-Rb})
\nonumber\\
\frac{\partial \bm{P^n}}{\partial t}=&&\gamma^n(\bm{B}+\lambda_{Rb-Ne}M_0^{Rb}\bm{P^e_{Rb}})\times \bm{P^n}+\bm{\Omega}\times \bm{P^n}
\nonumber\\
&&+R_{Rb-Ne}^{se}(\bm{P^e_{Rb}}-\bm{P^n})-\frac{1}{T_{2n},T_{2n},T_{1n}}\bm{P^n}
\end{eqnarray}
In the above equations, $\bm{P_{K}^e}$, $\bm{P_{Rb}^e}$ and $\bm{P^n}$ are respectively the polarization of K, Rb and $^{21}$Ne atoms. $\bm{\Omega}$ is the input rotation velocity. $Q(P^e_{K-Rb})$ is the slow down factor due to the rapid spin exchange\cite{allred2002}. Traditionally, due to the difference of the K and Rb nuclear spin, the slow down factor is different for K and Rb. However, the rapid spin exchange between the K and Rb spin can change the slow down factor of both K and Rb atoms. The slow down factors will be the same as they collision with each other. $\bm{B}$ is the external magnetic field. $\lambda_{K-Ne}M_0^n\bm{P^n}$ is a magnetic field which is produced by the $^{21}$Ne atoms through  spin exchange interaction\cite{Schaefer1989}. $\lambda_{K-Ne}$ is equal to $\frac{8}{3}\pi\kappa_{K-Ne}$ where $\kappa_{K-Ne}$ is a spin exchange enhancement factor due to the overlap of K electron wave function and the $^{21}$Ne nucleus. $M_0^n$ is equal to $\mu_{Ne}n_{Ne}$, where $\mu_{Ne}$is the nuclear magnetic moment and $n_{Ne}$ is the density of $^{21}$Ne atoms. $\bm{L^K}$ and $\bm{L^{Rb}}$ are the ac Stark shifts of the K atoms and Rb atoms. $R_p$ is the pumping rate of K atoms while $\bm{s_p}$ gives its direction and magnitude of the photon spin polarization. $R_{Rb-K}^{SE}$ is the spin exchange transfer rate from Rb atoms to K atoms. In K-Rb-$^{21}$Ne co-magnetometer, the spin relaxation rate of Rb atoms mainly comes from the spin destruction rate between the Rb and Rb atoms $R_{sd}$ and the spin exchange relaxation rate of Rb atoms $Q(P^e_{K-Rb})/{T_2^{SE}}$ which is caused by the Rb magnetization field\cite{fangjiancheng2016}. Due to the large density ratio between Rb and K, the magnetic field produced by the electron spins $\lambda_{Rb-Ne}M_0^{Rb}\bm{P^e_{Rb}}$ which is experienced by the $^{21}$Ne spins mainly comes from the Rb atoms spins. $R_{Rb-Ne}^{se}$ is the spin exchange transfer rate from Rb to $^{21}$Ne atoms. $1/T_{2n}$ and $1/T_{1n}$ are respectively transverse and longitude  relaxation rate of the $^{21}$Ne atomic spins.

After hours of spin exchange optical pumping of the $^{21}$Ne spins. The polarization of $^{21}$Ne is in its steady state. Suppose that $B_y$ magnetic field and $B_z$ magnetic fields are applied to the co-magnetometer. The ac Stark shifts of K and Rb are in the z direction. Suppose that small $B_y$ is applied to the co-magnetometer that the angle between the polarization of the spins and the z direction are small. The polarization of K, Rb and $^{21}$Ne in the z direction are constants. At the steady state, it is clear that $\frac{\partial \bm{P^e_K}}{\partial t}=0$, $\frac{\partial \bm{P^e_{Rb}}}{\partial t}=0$ and $\frac{\partial \bm{P^n}}{\partial t}=0$. The polarization in the z direction could be easily calculated to be:
\begin{eqnarray}\label{equation2}
&&P^e_{zK}\approx  P^e_{zRb}\approx\frac{D_r R_p}{D_r R_p+R_{sd}+Q(P^e_{K-Rb})/T_2^{SE}}\nonumber\\
&&(R_p\ll R_{Rb-K}^{SE}, D_r\ll 1)\nonumber\\
&&P^n_{z}\approx \frac{R_{Rb-Ne}^{se}}{R_{Rb-Ne}^{se}+1/T_{1n}}
\end{eqnarray}
As the density ratio of K to Rb is kept at a low level which is about 0.01 in our experiment to improve the uniformity of the Rb polarization\cite{smiciklas2010}. K and Rb atoms are in fast spin exchange and the exchange rate is at the order of 1$\times$10$^6$ sec$^{-1}$. The pumping rate of K $R_p$ is kept to be approximately $ (R_{sd}+Q(P^e_{K-Rb})/T_2^{SE})/D_r$ to achieve the condition of 50$\%$ polarization of Rb atoms. According to the conditions in a K-Rb-$^{21}$Ne co-magnetometer, $R_p$ is approximately on the order of 1$\times$10$^5$ sec$^{-1}$. According to the above discussion, the conditions in equation \ref{equation2} are realized.

A probe beam is used to detect the projection of electron spin polarization along the propagation direction(x direction). Rb D1 or Rb lines are utilized to detect the polarization. The polarization of K and Rb in the x direction could be solved and is given by:
\begin{eqnarray}\label{equation3}
&&P^e_{xK}\approx  P^e_{xRb}\nonumber\\
&&\approx P^e_{zRb}\frac{R_{tot}^{Rb}/\gamma^e (\frac{B_y}{B^n}\delta B_z+\frac{\Omega_y}{\gamma^n}+\frac{(D_r L_z^K+L_z^{Rb})}{R_{tot}^{Rb}/\gamma^e})\frac{\Omega_x}{\gamma^n}}{(R_{tot}^{Rb}/\gamma^e)^2+(\delta B_z+D_r L_z^K+L_z^{Rb})^2}\nonumber\\
&&(where, R_{tot}^{Rb}=D_r R_p+R_{sd}+Q(P^e_{K-Rb})/T_2^{SE},
\nonumber\\
&&\delta B_z=B_z-\lambda_{Rb-Ne}M_0^{Rb} P^e_{zRb}-\lambda_{Rb-Ne}M_0^n P^n_z,\nonumber\\
&&B^n=\lambda_{Rb-Ne}M_0^n P^n_z)
\nonumber\\
&&(R_p\ll R_{Rb-K}^{SE}, D_r\ll 1)
\end{eqnarray}
The polarization of Rb is measured in the K-Rb-$^{21}$Ne co-magnetometer. There is a total ac Stark shift $D_r L_z^K+L_z^{Rb}$ in equation \ref{equation3} as we measure polarization of Rb. The ac Stark shift of K $ L_z^K$ is suppressed by a factor $D_r$ will be measured by the Rb polarization. We say that the light shifts of K and Rb are mixed by the spin exchange interaction. A $B_y$ modulation method could be utilized to measure the mixed ac Stark shifts\cite{fangjiancheng2016}. A square $B_y$ magnetic field is applied to the co-magnetometer and the steady state square wave response amplitudes are measured as $\delta B_z$ is changed. The details of the measurement is shown in the next section.
\section{EXPERIMENTAL SETUP AND RESULTS}
\begin{figure}
\centering
\includegraphics[width=6cm,height=6cm]{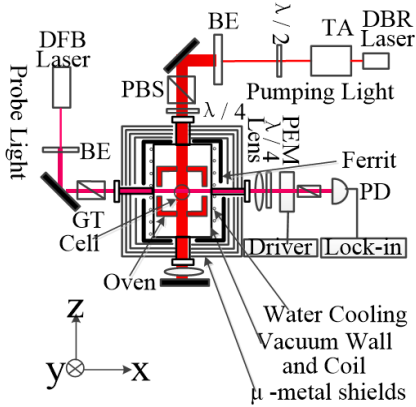}
\caption{\label{figure1} The schematic of the experimental setup.}
\end{figure}

A K-Rb-$^{21}$Ne co-magnetometer is constructed to measure the mixing of the K and Rb ac Stark shifts. The apparatus is similar to the one shown in this reference\cite{fangjiancheng2016}. We just show some main features of the apparatus. The co-magnetometer consists of a 1.4cm diameter spherical aluminosilicate glass (Schott8436 glass) vapor cell that contains a small droplet of K-Rb ($^{85}$Rb (72.2$\%$) and $^{87}$Rb (27.8$\%$)) mixture, about 3.0Atm of $^{21}$Ne gas (70$\%$ isotope enriched) and 40 torr of N$_2$ gas for quenching. The cell is heated to be 473 K and the oven was in vacuum for thermal insulation. Water cooling system is utilized to protect the magnetic shields from the thermal radiation of the oven. $\mu$-metal magnetic field shields and ferrit are utilized to shield the earth magnetic field. The residual magnetic field in the shields is further compensated by a three axis coil. K atoms are directly polarized by pumping light whose wavelength is on D1 line of K atoms. The pump  distributed bragg reflector(DBR) laser is amplified by a taper amplifier(TA) to about 1W. The power density of the pumping light before the vapor cell is set to be about 620mW/cm$^2$. A polarization beam splitter(PBS) and a 1/4 wave plate are combined to change the pump light to circular polarized photons. The probe distributed feedback(DFB) laser is tuned about 0.3 nm away from the absorption center of the Rb D1 line. Both of the vertical pump beam and the horizon probe beam are expanded by beam expanders(BE). The photo elastic modulation (PEM) technique with two high distinction ratio Glan-Thompson polarizers(GT) is utilized to measure the rotation of the polarization plane of the linear polarized light. The laser intensity is accepted by the photo detector(PD) and amplified by the Lock-in amplifier.  The schematic of the experimental setup is shown in figure \ref{figure1}

\begin{figure}
\centering
\includegraphics[width=8cm,height=6cm]{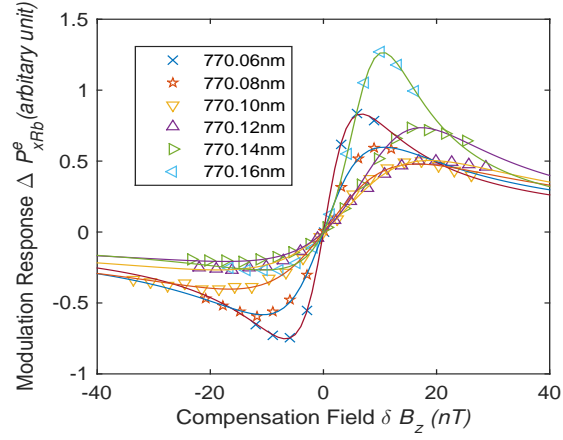}
\caption{\label{figure2} The relationship between the compensation field and the modulation response for different pumping wavelength. The measured data are fitted to equation \ref{equation4} to get the total ac Stark shifts.}
\end{figure}

We can measure the mixed ac Stark shifts $D_r L_z^K+L_z^{Rb}$ in equation \ref{equation3} in the co-magnetometer. Suppose that a step magnetic field $\Delta B_y$ in the y direction is applied to the co-magnetometer. Then there will be a step signal of $\Delta P^e_{xRb}$ which is directly detected by the probe laser. We take the derivative of $P^e_{xRb}$ in equation \ref{equation3} and we can get:
\begin{eqnarray}\label{equation4}
\frac{\Delta P^e_{xRb}}{\Delta B_y}
 = \frac{P^e_{zRb} R_{tot}^{Rb}/\gamma^e}{(R_{tot}^{Rb}/\gamma^e)^2+(\delta B_z+D_r L_z^K+L_z^{Rb})^2} \frac{\delta B_z}{B^n}
\end{eqnarray}
As a square magnetic field $\Delta B_y$ is applied to the co-magnetometer, the steady state output response $\Delta P^e_{xRb}$ is measured. The above measurement is repeated as the magnetic field in the z direction $B_z$ is changed. As the compensation magnetic field $\delta B_z$ is changed, according to equation \ref{equation4}, the response of the Rb polarization in x direction $\Delta P^e_{xRb}$ changes as a Lorentz function. Due to the existence of the ac Stark shifts, the symmetry Lorentz function changes into asymmetry one. FIG. \ref{figure2} shows the measured relationship between the compensation field and the modulation response. As we change the wavelength of the pump laser, the symmetry of the Lorentz function change. The fitted results could be used to get the total ac Stark shift $D_r L^K_z+L^{Rb}_z$. We just show parts of the measured data for it is hard to see the results clearly if all the data is plot in one figure.

\begin{figure}
\centering
\includegraphics[width=8cm,height=6cm]{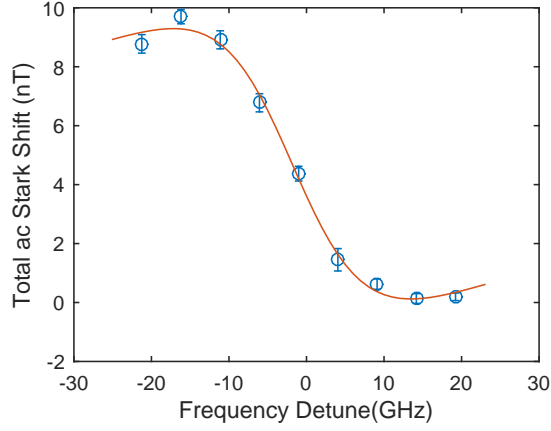}
\caption{\label{figure3} The relationship between pumping laser frequencies and the total ac Stark shifts which are measured by the co-magnetometer.}
\end{figure}

FIG. \ref{figure3} shows the relationship between the total ac Stark shift and the pump laser frequency detune. The x axis unit is changed from nm into GHz. The frequency at 0 GHz is at the 770.108nm which is the absorption peak of the D1 line of K atoms as there is no broadening of the line. As we change the frequency of the pumping laser, the ac Stark shifts of K atoms is due to virtual transitions\cite{Appelt1998} and it is given by:
\begin{eqnarray}\label{equation5}
L_z^K=\frac{\Phi_{D1}^K r_e c f_{D1}^K}{A\gamma^e}\frac{(\Gamma_{D1}^K/2)}{((\nu-\nu_{0D1}^K)^2-(\Gamma_{D1}^K/2))^2}
\end{eqnarray}
where $\Phi_{D1}^K$ is the photon number flux and $A$ is the transverse area of the pumping light. $r_e$ is the electron radius and $c$ is the velocity of light. $f_{D1}^K$ is the oscillator strength\cite{Caliebe1979} and $\gamma^e$ is the electron gyro-magnetic ratio. $\Gamma_{D1}^K$ is the line width of the pressure and power broadened absorption line of K D1 line with the 3Atm $^{21}$Ne atoms\cite{Lwin1978}. $\nu_{0D1}^K$ is the absorption center of the absorption line. The absorption center is not at 770.108nm and it is shifted by the collision with noble gas\cite{Lwin1978}. The far off-resonant pumping laser will cause ac Stark shift of Rb atoms on the D2 line. Due to the large detune of the laser which is much larger than the pressure broadened absorption line width of Rb-$^{21}$Ne pair, the ac Stark shift of Rb can be simplified to be:
\begin{eqnarray}\label{equation6}
L_z^{Rb}=\frac{\Phi_{D1}^K r_e c f_{D1}^{Rb}}{A\gamma^e}(\frac{1}{\nu-\nu_{0D2}^{Rb}}-\frac{1}{\nu-\nu_{0D1}^{Rb}})
\end{eqnarray}
As we change the pumping laser frequency, the ac Stark shift of Rb is approximately constant. Combining equation\ref{equation5} with equation\ref{equation6}, we fit the total ac Stark shift $D_r L_z^{K}+L_z^{Rb}$ with the measured data in FIG. \ref{figure3}. We can conclude that the total ac Stark shift which is measured by the Rb polarization in the x direction is the mixing of the Rb and K ac Stark shifts and the K ac Stark shift is suppressed by a factor of $D_r$.

\section{DISCUSSION}
The line width of the Lorentz function in FIG. \ref{figure3} is fitted to be 30.6GHz. Theoretical calculation result shows that the pressure broadened line width of K in 3.0Atm $^{21}$Ne is 23.4GHz\cite{Lwin1978}. The laser power used in this experiment is high. The power broadening of the absorption line can not be neglected. The total line width of the pressure broadening and power broadening\cite{Citron1977} is given by:
\begin{eqnarray}\label{equation7}
\Gamma_{D1}^K=\Gamma_{P-D1}^K(1+\frac{I}{I_{sat}})^{\frac{1}{2}}
\end{eqnarray}
where $\Gamma_{P-D1}^K$ is the broadening line width due to pressure broadening. $I_{sat}$ is the saturation intensity which is equal to $(\pi h c)/(3 \lambda ^3  \tau)$. Here $\tau=(\Gamma_{P-D1}^K)^{-1}$ is the lifetime for radiative broadening. The resonance transition of K is 770.108nm and the radiative broadening in the 3Atm vapor cell is 23.4GHz. These lead to an intensity $I_{sat}$=1.0W/cm$^2$. The power density is 620mW/cm$^2$ and the total broadening of the line is 29.4GHz. This is coincident with the fitted result of 30.6GHz. There is an constant 4.7nT in the fitting. The constant is the ac Stark shift of Rb. The ac Stark shift of Rb is calculated to be 4.5nT as the pumping power density is 620mW/cm$^2$. This is coincident with the measured result of 4.7nT. We can also get the  density ratio of K to Rb to be 0.0037. The ac Stark shift could be further suppressed by change the vapor cell to a larger density ratio of K to Rb.

The mixing of the ac Stark shift of two alkali metal ensembles is closing related to the density ratio of the two ensemble. If the density ratio is large, we need to tune the laser wavelength far away from the resonance center of the absorption line to compensate the ac Stark shift of the other ensemble. This could lead to reduction of pumping rate. One need to use vapor cells with small density ratio to avoid this problem. On the other hand, we could use two spin ensembles whose wavelengths are far away from each other. The Cs-Rb pair could be utilized to compensate the ac Stark shift. The pumping Cs D1 line is at 894nm which is far away from the 795nm Rb D1 line. The ac Stark shift which needs to be compensate is relatively small compared to the K-Rb pair. We only need to tune small wavelength away from the absorption center.
\section{CONCLUSION}
In conclusion, we find the mixing of K and Rb ac Stark shifts through rapid spin exchange collision. The spin exchange collision between them rapidly change the direction of the spin and conserve the total angular momentum. The K ac Stark shift is suppressed by a factor of $D_r$ which is the density ratio between K and Rb. As ac Stark shift is always a systematic error in precision measurement. With this finding, we can get rid of ac Stark shift by compensating the ac Stark shift of one alkali atom with the other.
\section{ACKNOWLEDGMENTS}
This work is supported by the Key Programs of National Natural Science Foundation of China under Grant No. 61227902, 61374210 and 863 Plan of China.

\end{document}